\documentstyle[twocolumn,aps,epsf]{revtex}

\begin{document}
\draft
\title { Directed Flow of $\Lambda$-Hyperons in 2~-~6~AGeV Au~+~Au Collisions  }
\author{P.~Chung$^{(1)}$, N.~N.~Ajitanand$^{(1)}$,
J.~M.~Alexander$^{(1)}$,
M.~Anderson$^{(5)}$, D.~Best$^{(3)}$,F.P.~Brady$^{(5)}$,
T.~Case$^{(3)}$,
W.~Caskey$^{(5)}$, D.~Cebra$^{(5)}$,J.L.~Chance$^{(5)}$,
B.~Cole$^{(10)}$,
K.~Crowe$^{(3)}$, A.~Das$^{(2)}$, J.E.~Draper$^{(5)}$,
M.L.~Gilkes$^{(1)}$,
S.~Gushue$^{(1,8)}$, M.~Heffner$^{(5)}$,A.S.~Hirsch$^{(6)}$,
E.L.~Hjort$^{(6)}$, L.~Huo$^{(12)}$, M.~Justice$^{(4)}$,
M.~Kaplan$^{(7)}$, D.~Keane$^{(4)}$, J.C.~Kintner$^{(11)}$,
J.~Klay$^{(5)}$,D.~Krofcheck$^{(9)}$,R.~A.~Lacey$^{(1)}$,
J.~Lauret$^{(1)}$, M.A.~Lisa$^{(2)}$,H.~Liu$^{(4)}$, Y.M.~Liu$^{(12)}$,
R.~McGrath$^{(1)}$, Z.~Milosevich$^{(7)}$,G.~Odyniec$^{(3)}$,
D.L.~Olson$^{(3)}$,S.~Y.~Panitkin$^{(4)}$, C.~Pinkenburg$^{(1)}$,
N.T.~Porile$^{(6)}$,G.~Rai$^{(3)}$, H.G.~Ritter$^{(3)}$,
J.L.~Romero$^{(5)}$,
R.~Scharenberg$^{(6)}$, L.~Schroeder$^{(3)}$,B.~Srivastava$^{(6)}$,
N.T.B~Stone$^{(3)}$, T.J.M.~Symons$^{(3)}$, T.~Wienold$^{(3)}$,
R.~Witt$^{(4)}$ J.~Whitfield$^{(7)}$,
L.~Wood$^{(5)}$, and W.N.~Zhang$^{(12)}$
                       \\  (E895 Collaboration) }
        \address{$^{(1)}$Depts. of Chemistry and Physics, SUNY
          Stony Brook, New York 11794-3400
\\
$^{(2)}$Ohio State University, Columbus, Ohio 43210
\\
$^{(3)}$Lawrence Berkeley National Laboratory,Berkeley, California,
94720
\\
$^{(4)}$Kent State University, Kent, Ohio 44242
\\
$^{(5)}$University of California, Davis, California, 95616
\\
$^{(6)}$Purdue University, West Lafayette, Indiana, 47907-1396
\\
$^{(7)}$Carnegie Mellon University, Pittsburgh, Pennsylvania 15213
\\
$^{(8)}$Brookhaven National Laboratory, Upton, New York 11973
\\
$^{(9)}$University of Auckland, Auckland, New Zealand
\\
$^{(10)}$Columbia University, New York, New York 10027
\\
$^{(11)}$St. Mary's College, Moraga, California  94575
\\
$^{(12)}$Harbin Institute of Technology, Harbin, 150001 P.~R. China
\\
}

\date{\today}
\maketitle
\newpage
\begin{abstract}
%

        Directed flow measurements for $\Lambda$~-~hyperons
are presented and compared to those for protons produced in the
same Au~+~Au collisions (2, 4, and 6~AGeV; $b < 5 - 6$~fm).
The measurements indicate that $\Lambda$~-~hyperons flow consistently
in the same direction and with smaller
magnitudes than those of protons.  Such a strong positive flow
[for $\Lambda$s] has been predicted in calculations which include
the influence of the $\Lambda$-nucleon potential. The experimental flow
ratio
$\Lambda$/p is in qualitative agreement with expectations ($\sim 2/3$)
from the quark counting rule at 2 AGeV but is found to decrease with
increasing beam energy.

\end{abstract}

\narrowtext

	A pervasive theme of current relativistic heavy ion
research is the creation and study of nuclear matter at high
energy densities\cite{Stocker86,Qmxx,Reisdorf97,Bass98}.
Central to these studies are questions related to how hadronic properties
may be modified in a hot and dense nuclear
medium\cite{GBrown91,Senger98}. Modifications to the
properties of $\Lambda$'s and kaons which influence their production and
propagation are of particular current interest
\cite{GBrown91,Kaplan86,Wass96,GQli96,Bratkovskaya97,Lutz98,GQli98,Wang98}.
This is because they have not only been linked to a chirally restored
phase of nuclear matter\cite{GBrown91},
but also to the detailed characteristics of the high density nuclear
material existing in neutron stars\cite{GQli97,Thorsson94}.

	Recent theoretical studies of the propagation of $\Lambda$'s and
kaons, in hot and dense nuclear matter, have identifed characteristic
flow patterns for these particles which could serve as an important
probe for their respective in-medium
potentials\cite{GQli96,Bratkovskaya97,GQli98,Wang98}.
Subsequently, results from kaon measurements have
been attributed to the influence of the vector component of the
in-medium kaon-nucleon potential\cite{Ritman95,Yshin98,PChung98}.
The propagation of $\Lambda$'s [in nuclear matter] is predicted to
differ from that of kaons due to these different interactions.
For example, at relatively
low energies the $\Lambda$-nucleon scattering cross section is much
larger than
the kaon-nucleon cross section. More importantly, the kaon potential
is thought to be weakly repulsive while that for $\Lambda$-hyperons is
believed to be attractive for the conditions expected in
1-10~AGeV heavy ion collisions\cite{GQli98}.
One of the possible manifestations of these differences is a
predicted pattern for directed flow of $\Lambda$'s which is distinguishable
from that for kaons and protons\cite{GQli96,GQli98,Wang98}. Theoretical
studies of the directed flow of $\Lambda$'s show that
the $\Lambda$ flow is relatively insensitive to the magnitude
of the $\Lambda$-nucleon cross section, but has greater
sensitivity to the $\Lambda$-nucleon potential\cite{GQli96,GQli98}.
Thus,
it is important to obtain a set of $\Lambda$ flow
measurements in the 1-10 AGeV beam energy range and to test
its utility as an important constraint for the $\Lambda$-nucleon
potential.

	In this letter, we present the first experimental flow excitation
function for $\Lambda$~-~hyperons at AGS energies,
and compare it to that for protons produced in the same Au + Au
collisions. These measurements are important because
they provide a unique opportunity for probing the $\Lambda$-nucleon
potential in collisions which are predicted to produce the highest
baryon densities\cite{Pang92}. Such an opportunity is not afforded
by hypernuclei studies which allow the investigation of the
$\Lambda$-nucleon potential at or below normal nuclear matter density.
Knowledge of the $\Lambda$~-~nucleon potential at high baryon densities
is crucial for accurate predictions of the amount of strangeness-bearing
matter in the interior of neutron stars\cite{prakash98}

        Measurements have been performed with the E895 detector
system at the Alternating Gradient Synchrotron~ at the
Brookhaven National Laboratory. Details on the detector and its setup have been
reported earlier\cite{PChung98,GRai90,CPinkenburg99}. Suffice it to say,
the data presented here benefit from the excellent coverage,
continuous 3D-tracking, and particle
identification
capabilities of the TPC.
These features are crucial for the efficient detection
and reconstruction of $\Lambda$'s and for accurate flow determinations.

        The $\Lambda$-hyperons have been reconstructed from the
daughters
of their charged particle decay,
$\Lambda \longrightarrow p + \pi^-$ (branching
ratio $\sim$ 64\%) following the procedure outlined in
Refs.~\cite{PChung98}~and~\cite{Justice98}.
All TPC tracks in an event were reconstructed followed by the
calculation of an overall event vertex. Thereafter, each $p\pi^-$
pair was considered and its point of closest approach obtained.
Pairs whose trajectories intersect (with fairly loose criteria such as decay
distance from event vertex  $>$ 0.5 cm ) at a point
other than the main event vertex were assigned as $\Lambda$ candidates
and evaluated to yield an invariant mass and associated momentum.
These $\Lambda$ candidates were then passed
to a fully connected feedforward multilayered neural
network\cite{Justice97}
trained to separate ``true'' $\Lambda$'s from the combinatoric background.
The
network was trained from a set consisting of ``true'' $\Lambda$'s and a
set
consisting of a combinatoric background. ``True'' $\Lambda$'s were
generated by
tagging and embedding simulated $\Lambda$'s in raw data events in
a detailed GEANT simulation of the TPC. The combinatoric background
or ``fake'' $\Lambda$'s were generated via a mixed event procedure in
which the daughter particles of the $\Lambda$ ($p \pi^-$) were chosen
from
different data events.

\begin{figure}
\centerline{\epsfysize=4.5in \epsffile{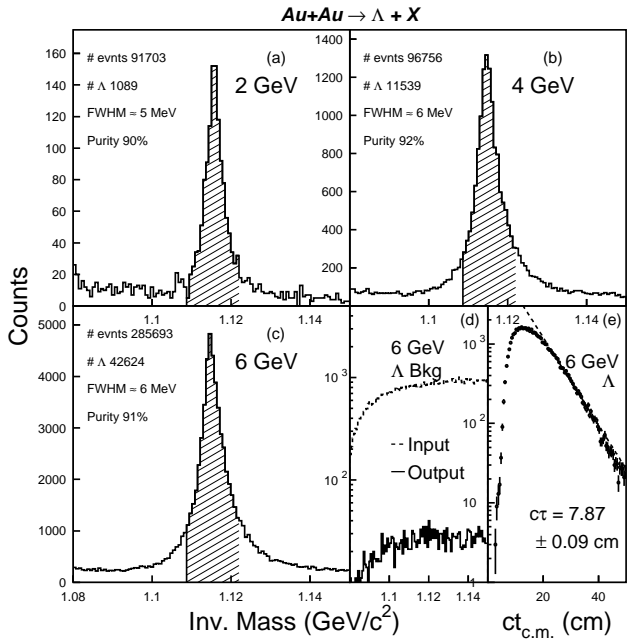}}
\vspace*{-.5in}
\caption{   Invariant mass distribution for $\Lambda$ hyperons measured
at 2, 4, 6 AGeV as indicated (b $<$ 6fm). Panel (d) shows input (dashed curve)
and output (solid curve) invariant mass distributions for combinatoric events.
Panel (e) shows the decay length distribution.
}
\label{lam_mass_qua}
\end{figure}

        The invariant mass distributions for $\Lambda$'s
obtained from the neural network are shown in
Figs.~\ref{lam_mass_qua}a~-~\ref{lam_mass_qua}c
for 2, 4, and 6~AGeV Au~+~Au collisions respectively\cite{enote}.
It is noteworthy that we have verified that the procedure used to train
the
neural network does not lead to the spurious ``creation''
of $\Lambda$'s (see Ref.~\cite{PChung98}). That is, combinatoric
background processed through the neural network does not lead
to $\Lambda$ peaks in the invariant-mass distribution as demonstrated
for
6~AGeV data in Fig.~\ref{lam_mass_qua}d.

The distributions shown in Fig.~1 have been obtained for central and mid-central
events in which one or more $\Lambda$'s have been detected. Using the
charged particle multiplicity of an event as a measure of its centrality,
 we estimate that these events
are associated with an impact-parameter range b~$\lesssim 5-6$~fm.
The distributions shown in
Figs.~\ref{lam_mass_qua}a~-~\ref{lam_mass_qua}c
show relatively narrow invariant mass peaks (Full Width at Half Maximun
 $\sim 6$~MeV)
at the characteristic value expected for the $\Lambda$ hyperon
($\sim 1.116$~GeV). The distributions also show an excellent peak
to background ratio which clearly attests to the reliability of the separation
from combinatoric background.

Fig.~\ref{lam_mass_qua}e shows
a typical decay-length distribution (in the c.m. frame) for
$\Lambda$'s obtained at 6~AGeV. A deficit below
ct~$\sim$~20 cm reflects the difficulty of $\Lambda$ reconstruction
in the region of high track density near to the main event vertex.
Deficiencies above
40 cm reflect inefficiencies associated with $\Lambda$ decays
close to the edge of
the TPC. An exponential fit to the distribution
over a region for which the detection efficiency is constant
($\sim 22~-~38$~cm), yields a c$\tau$ value of $7.9\pm 0.1$~cm ($\chi^2 \sim
0.9$). This value is close to the expected value of 7.8 cm.

\begin{figure}
\centerline{\epsfysize=4.in \epsffile{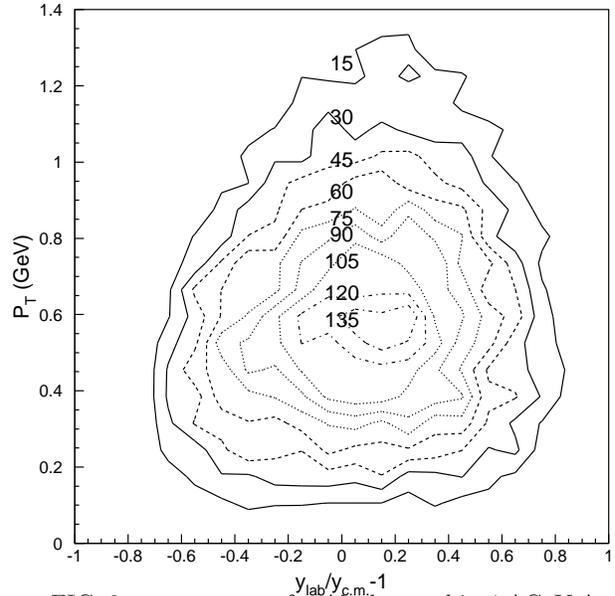}}
\vspace*{-.5in}
\caption{$<p_T>$ vs $y_0$  for $\Lambda$'s detected in 4 AGeV Au~+~Au collisions.
Contour levels are indicated.
}
\label{4GEV_LAM_PT_YCM}
\end{figure}

A better appreciation of the TPC coverage can be obtained from the $P_t$ vs rapidity plot
(cf Fig.\ref{4GEV_LAM_PT_YCM}) obtained for 4 AGeV
$\Lambda$'s. One can identify 2 areas of low acceptance : (1)  $P_T <$ 100 MeV
, due to the high track density in a cone around the beam,
(2) rapidity $<$ -0.6 , due to the finite volume of the TPC. These two types
of losses exhibit different trends
with beam energy. The first one is least prominent at 2 AGeV
and becomes gradually worse at 6 AGeV due to the higher track density. The
second one is worst at 2 AGeV but improves gradually with
increasing beam energy as a result of the increasing nucleon-nucleon c.m.
rapidity with beam energy.

	The method for analysis of the sideward flow of protons has
been discussed in a prior publication\cite{HLiu99}.
Here we focus only on the procedure employed for $\Lambda$'s.
The hatched area centered on the invariant mass peaks shown
in Figs.~\ref{lam_mass_qua}a~-~\ref{lam_mass_qua}c
($1.11 \le m_{inv} \le 1.122$) represents the
mass gates used for the flow analysis. These gates
ensure a relatively pure  sample
of $\Lambda$'s ($\sim$ 90\%) for all beam energies.
On the other hand, this sample may include
secondary $\Lambda$'s which result from the decay of $\Sigma^0$'s.
These secondary $\Lambda$'s are experimentally indistinguishable
[via our analysis] from the primary ones but it is believed
that they do not constitute the bulk of the detected $\Lambda$'s.
Results from the Relativistic
Quantum Molecular Dynamics model\cite{Sorge95} indicate;
(a) a maximum ratio $\Sigma^0$/$\Lambda$, of $\sim 0.3$ for
the 2~-~6 AGeV Au~+~Au collisions relevant to our analysis and
(b) a $\Sigma^0$ flow magnitude which is similar to that
for $\Lambda$'s.

	The reaction plane was determined via the standard transverse
momentum analysis method of Danielewicz and Odyniec\cite{daniel85}
as described in detail in Ref.~\cite{CPinkenburg99}.
We note here that the daughter protons of $\Lambda$ candidates were
excluded
from the proton sample used for reaction plane determination.
Such a procedure eliminates any autocorrelation.
Deficiencies in the acceptance of the TPC result in a non-uniform
reaction plane distribution for all beam energies. We account for
such non-uniformities by applying rapidity and multiplicity dependent
corrections\cite{CPinkenburg99} which serve to flatten the reaction
plane distribution. The dispersion of the reaction plane was estimated
for 2,4 and 6 AGeV via the
the sub-event method\cite{daniel85}; values of the dispersion correction
are 1.08, 1.20 and 1.44 respectively\cite{CPinkenburg99}.

\begin{figure}
\centerline{\epsfysize=4.5in \epsffile{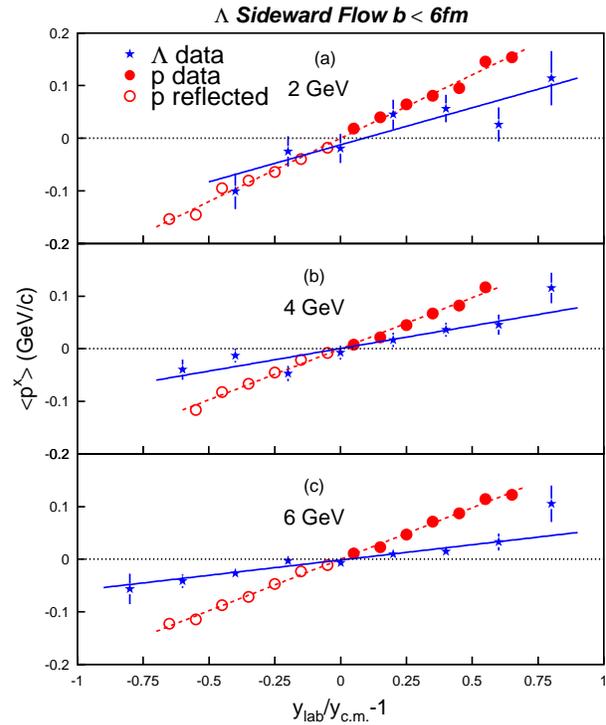}}
\vspace*{-.4in}
\caption{   $<p^x>$ vs $y_0$  for $\Lambda$'s
   and protons (b $<$ 6fm). The open circles indicate the reflected
   values for protons, and the solid and dashed-curves represent
   fits to the data. The $<p^x>$ values are corrected for reaction
   plane dispersion.
   }
\label{lam_flow_dou}
\end{figure}

	Figs.~\ref{lam_flow_dou}a~-~c show
representative results for the mean transverse momenta in the
reaction plane $<p^x>$, vs. the normalized
c.m. rapidity $y_0$, for protons and $\Lambda$'s
produced in the same events. $y_0 = y_{Lab}/y_{cm}
-1$,
where $y_{Lab}$ and $y_{cm}$ represent the rapidity of the emitted
particle in the Lab and the rapidity of the c.m respectively.
The $<p^x>$ values shown in Fig.~\ref{lam_flow_dou}
have been corrected for reaction
plane dispersion\cite{dan87,oll97,Postkanzer98} by the above 
multiplicative correction factors. The
values for the $\Lambda$'s also take account of a small correction
associated with the $\sim$~9\% combinatoric background at each
beam energy. The latter correction was made by evaluating
the $<p^x>$ for the experimental combinatoric background
[for each of several rapidity bins for each beam energy] followed
by a weighted subtraction of these values from the $<p^x>$ values
obtained for the invariant mass
selections indicated in Fig.~\ref{lam_mass_qua} (The
background flow is consistent with expectations for the uncorrelated
proton pion pairs).

Figs.~\ref{lam_flow_dou}a~-~c show
trends which clearly indicate that $\Lambda$'s and protons
flow consistently in the same direction. However, the magnitude of the
$\Lambda$ flow is consistently smaller than that for protons. 
These results are consistent with
prior $\Lambda$~-~flow measurements\cite{Ritman95,Justice98}
performed for beam energies $\lesssim$~2 A~GeV. However,
they are in stark contrast to the anti-flow pattern recently observed
for $K^0_s$ mesons at 6~A~GeV\cite{PChung98}.
This contrast is suggestive of differences in the kaon-nucleon
and $\Lambda$-nucleon potentials. Namely,
for a similar range of densities and momenta, the predicted
kaon-nucleon potential is repulsive while that for the
$\Lambda$-nucleon is attractive\cite{GQli96,GQli98,Wang98}.

\begin{figure}
\centerline{\epsfysize=4.5in \epsffile{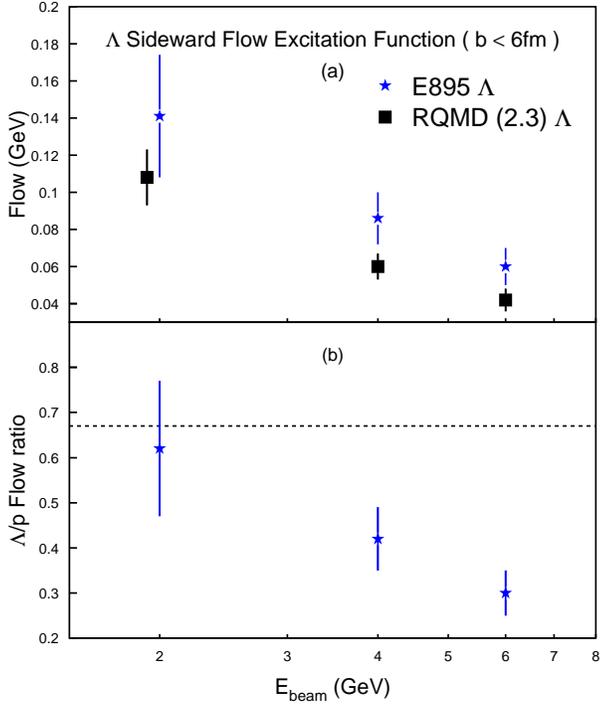}}
\vspace*{-.4in}
\caption{    (a) $\Lambda$ hyperon flow vs beam energy. Filled stars and
        squares
        represent data and RQMD results respectively. (b) $\Lambda$/p
	flow ratio vs beam energy for the same impact-parameter range.
	The dashed line indicates a value of 2/3.}
\label{lam_ex_fucn}
\end{figure}

In order to quantify the flow magnitude we have evaluated the slope
at mid-rapidity for protons $F_p$ and $\Lambda$'s $F_{\Lambda}$
($F_{p,\Lambda} = \frac{d<p^x>}{dy_0}|_{(y_0\sim 0)}$) of the
transverse
 momentum data (cf.~Figs.~\ref{lam_flow_dou}) .
The results obtained from linear fits to these data are summarized
in Figs.~\ref{lam_ex_fucn}a and~\ref{lam_ex_fucn}b.
Experimental and calculated $\Lambda$ flow excitation functions are
shown in
Fig.~\ref{lam_ex_fucn}a; the data and the results from
RQMD (v 2.3))\cite{Sorge95} calculations both indicate
a continuous decrease in the flow ($F$) with increasing
beam energy. However, the results from the calculation systematically
underpredict the experimentally observed flow magnitude. Since
rescattering effects are included in the RQMD calculations, it is
suggested that the difference may be due
to the absence of the $\Lambda$-nucleon potential in RQMD.
It is important to point out here that stronger $\Lambda$ flow has
been found in calculations which include the explicit effect of
the $\Lambda$ nucleon potential\cite{GQli96,GQli98,Wang98}.

The observed trend of the $\Lambda$ flow
is very similar to that observed for protons co-produced
with $\Lambda$'s, as well as for protons in which no
explicit condition for $\Lambda$ detection was imposed\cite{HLiu99}.
The decrease of $F$ with increasing beam energy is consistent with the
notion that the flow of primordial $\Lambda$'s reflects the
collective flow of the baryon-baryon and pion-baryon pairs from
which they are produced. This trend could also result from a weakening
of rescattering effects and/or the attractive $\Lambda$-nucleon potential.
Such a weakening of the $\Lambda$-nucleon potential has been predicted
for relatively large baryon densities\cite{GQli98}.

The quark counting rule asserts that $\Lambda$'s interact with nucleons only
through their non-strange quark constituents\cite{mosz73}. Since a $\Lambda$ particle has
only two such quarks, this suggests that the
$\Lambda$ potential is $\sim 2/3$ of that for nucleons\cite{mosz73}.
The experimental flow ratio $F_\Lambda / F_p$, is shown as a function of
beam energy in Fig.~\ref{lam_ex_fucn}b. The data indicate a value
of $\sim 2/3$ at 2 AGeV. This value is similar to that obtained for the
Ni~+~Cu system\cite{Justice98}, and is in qualitative agreement with
the suggested ratio of the $\Lambda$/p potential. On the other
hand, there is an unmistakeable deviation from the value 2/3
(indicated by the dashed line) with
increasing beam energy. Such a deviation suggests that a simple
scaling by 2/3 to obtain the $\Lambda$-nucleon potential may be an
oversimplification. The deviation could
be related to rescattering effects or to the detailed characteristics of
the $\Lambda$-nucleon potential for the densities produced
at 4 and 6~A~GeV. More detailed calculations are required to distinguish
between these two effects.


We have measured directed flow excitation functions
for $\Lambda$-hyperons and protons produced in the same
mid-central Au+Au collisions.  The data show positive flow for
$\Lambda$'s and protons for all beam energies. 
This observation is in stark contrast to the prominent
anti-flow observed for neutral kaons from
the same data set\cite{PChung98}.
This contrast is suggestive of the expected differences in the
kaon-nucleon and $\Lambda$-nucleon interactions for the densities
attained in 2~-~6 AGeV Au~+~Au collisions.
A comparison of the $\Lambda$ flow excitation function
to that obtained from RQMD, shows consistently larger experimental
values, suggesting an influence of the $\Lambda$-nucleon
interaction.
Both $\Lambda$ and proton flow show a decrease with increasing
beam energy. However, the flow ratio
$F_\Lambda /F_p$, decreases from $\sim 2/3$ at 2 AGeV to
$\sim 1/3$ at 6 AGeV. Such a trend is qualitatively inconsistent
with the quark counting rule and could be related to the detailed
features of the $\Lambda$-nucleon potential.
More studies of strange particle dynamics can illuminate additional 
properties of high density matter which is thought to be
rich in strangeness.


        This work was supported in part by the U.S. Department
of Energy under Grant No. DE-FGO2-87ER40331.A008 and other grants acknowledged
in Ref.\cite{CPinkenburg99}. We acknowledge fruitful discussions with M. Prakash.



%

%



\begin{references}
\bibitem{enote} Actual beam energies were 1.85, 3.9 and 5.9 AGeV.
 Measurements were also made at 7.9 AGeV but these results suffer
from rather poor statistics.
\end{references}
\end{document}